\documentclass{article}

\usepackage{arxiv}

\usepackage[utf8]{inputenc} 
\usepackage[T1]{fontenc}    
\usepackage{hyperref}       
\usepackage{url}            
\usepackage{booktabs}       
\usepackage{amsfonts}       
\usepackage{nicefrac}       
\usepackage{microtype}      
\usepackage{lipsum}
\usepackage{graphicx}
\usepackage{amsmath}
\graphicspath{ {./images/} }

\usepackage{url}

\title{Speed-up Heuristic for an On-Demand Ride-Pooling Algorithm}

\author{
  Roman Engelhardt \\
  Chair of Traffic Engineering and Control \\
  Technical University of Munich\\
  \texttt{roman.engelhardt@tum.de} \\
   \And
  Florian Dandl \\
  Chair of Traffic Engineering and Control \\
  Technical University of Munich\\
  \texttt{florian.dandl@tum.de} \\
  \And
  Klaus Bogenberger \\
  Chair of Traffic Engineering and Control \\
  Technical University of Munich\\
  \texttt{klaus.bogenberger@tum.de} \\
}

\begin{document}
\maketitle



\begin{abstract}
With ongoing developments in digitalization and advances in the field of autonomous driving, on-demand ride pooling is a mobility service with the potential to disrupt the urban mobility market. Nevertheless, to apply this kind of service successfully efficient algorithms have to be implemented for effective fleet management to exploit the benefits associated with this mobility service. Especially real time computation of finding beneficial assignments is a problem not solved for large problem sizes until today. In this study, we show the importance of using advanced algorithms by comparing a fast, but simple insertion heuristic algorithm with a state-of-the-art multi-step matching algorithm. We test the algorithms in various scenarios based on private vehicle trip OD-data for Munich, Germany. Results indicate that in the tested scenarios by using the multi-step algorithm up to 8$\%$ additional requests could be served while also 10$\%$ additional driven distance could be saved. However, computational time for finding optimal assignments in the advanced algorithm exceeds real time rather fast as problem size increases. Therefore, several aspects to reduce the computational time by decreasing redundant checks of the advanced multi step algorithm are introduced. Finally, a refined vehicle selection heuristic based on three rules is presented to furthermore reduce the computational effort. In the tested scenarios this heuristic can speed up the most cost intensive algorithm step by a factor of over 8, while keeping the number of served requests almost constant and maintaining around 70$\%$ of the driven distance saved in the system. Considering all algorithm steps, an overall speed up of 2.5 could be achieved.
\end{abstract}


\section{Introduction}
The global growth of population and the resulting urbanization stretches the limits of cities' resources and infrastructure all over the world. Rising demand on transportation tends to congest street networks for an increasing fraction of the day, leading to waste of energy, air pollution, economic costs and personal stress for vehicles' drivers. Especially private vehicle trips have a big influence on the exceeded network capacities due to their highly ineffective way of transport induced by low average occupancy and high space consumption.
\par
With recent developments in digitalization and communication on-demand mobility systems (ODM) started to emerge rapidly with no end of growth in sight \cite{growthODMchina}. Among many subgroups of ODM-systems like car-sharing or taxi-like systems, On-Demand Ride-Pooling (ODRP) is a promising candidate to improve conditions in urban traffic. In this mobility service an operator matches customers with similar travel requests, based on travel origins, destinations and points in time, whereby these customers can share parts of their trips leading to higher vehicle occupancy and therefore increased effectivity of transport.
\par
Recent theoretical studies regarding ride-pooling showed, that the probability of matching two customers exponentially approaches one with increasing demand in a ride-pooling system ~\cite{Tachet2017Scaling}. Because, also the prices for a trip can be shared among customers, a successful implementation may lead to a positive feedback reaction between price and demand, especially once the cost for an ODRP trip undercuts the costs for a trip with the private vehicle ~\cite{HERMINGHAUS201915}. Additionally, the introduction of autonomous driving may act as a catalyst for ODRP systems, because the largest cost component, the driver, will be removed from the business model.
\par
In order to provide such a mobility service an operator has to implement an efficient algorithm to match customers and assign resulting trips to its operating vehicles in a best possible way to optimize its objectives like cost reduction or reduced customer inconveniences. Besides effective fleet management, replies to customers' requests have to be provided in short terms to satisfy the customers' needs. However, since Ride-Pooling can be formulated as a dynamic vehicle routing problem \cite{Psaraftis1995}, the NP-hard formulation of the problem results in exploding computational time even for medium system sizes, preventing precise short-term replies based on good solutions. The curse of dimensionality becomes a problem even for single-occupancy algorithms when coordinated routing is considered \cite{LiangAlmeida2018}. For ride-pooling, the theoretical amount of possibilities to combine multiple requests into tours of multiple vehicles grows exponentially even with pre-processed routes between all access points of users: $O(m n^\mu)$, where $m$ is the number of vehicles, $n$ the number of requests and $\mu$ the maximum amount of requests per vehicle \cite{alonsomora}. Among the simplest known approaches to ride-pooling are rule-based insertion heuristics \cite{JungAnnealing} \cite{FagnantKockelman2018} \cite{Bischoff2017}. Jung et al. \cite{JungAnnealing} illustrate the benefits of using hybrid simulated annealing over an insertion heuristic in a study where shared taxis serve up to 18k requests in a 4-hour period in real-time. Santi et al. \cite{SantiShareability} introduce the concept of vehicle share-ability networks, which translates the vehicle routing problem into a graph problem with efficient solution.
\par
Alonso-Mora et al. \cite{alonsomora} introduced a multi-step algorithm for the ODRP-problem on which this paper is built on. Making use of share-ability networks and hard time constraints in possible routes, the ODRP-problem can be decoupled efficiently and problem sizes of the magnitude of Manhattan's taxi demand could be simulated in reasonable time. However, computational time still increases rapidly especially as fleet size increases. For their study in Manhattan, NYC, Alonso-Mora et al. report exceeding real time on a 24-core computer with a fleet size of 3000 vehicles \cite{alonsomora_supp}. We also observe a similar value for our in the following described study in Munich, Germany without any heuristics.
\par
To overcome these limitations, we therefore describe in this work several methods to improve the algorithm proposed in ~\cite{alonsomora}. Among these aspects is search space pruning, which was also proposed by \cite{samitha2019}. Furthermore, we introduce a heuristic to limit the number of possible vehicles per request based on three rules to decrease the search space additionally. Finally, we compare the performance with a simple insertion heuristic to stress the importance of the matching algorithm for effective fleet management.
\par
The rest of the paper is structured as follows: In section \ref{sec:methods} we describe the simulation setup and the algorithm, in section \ref{sec:caseStudy} we present comparisons of performance for the different algorithms for our case study of Munich, Germany. Finally, section \ref{sec:conclusion} concludes this study.

\section{Research Methodology}
\label{sec:methods}
In this section, we explain the basic setup of the agent-based simulation framework that we developed to model the matching of on-demand requests and vehicles. We start describing the basic simulation setup and the customer model defining the constraints of the allowed tours for the matching algorithms. Finally, the different fleet operator policies and matching algorithms are explained in detail.

\subsection{Basic Setup}
In this study, we model customers, vehicles and a fleet operator, as agents. They interact on a street network graph $G=(N,E)$, where $N$ are nodes and $E$ are edges of the street network. An edge $e \in E$ has a time-dependent travel time $t_e (t)$, which is used by the operator for route calculations. Moreover, all vehicles require the same $t_e (t)$ to travel through this edge. $N_a \subset N$ is a set of access points, where customers can start and end their trip.
\par
We define a route as a sequence of elements of $G$ combining an origin $o \in G$ and a destination $d \in G$. Furthermore, we define a tour as a sequence of routes that can be separated by stops at access points, where customers can board or leave a vehicle. Boarding and disembarking require the vehicle to stop for $T^B$, the boarding time, at an access point.
\par
The simulation runs for 24 hours and every time step $\Delta t$, the status of the fleet of vehicles is updated, i.e. vehicles with assigned tours either move closer to their next stop or customers board or leave the vehicle. In this model, the fleet operator does not make the decision to re-assign tours to its vehicles every time step $\Delta t$; instead decisions are made every $\Delta T^{dec}$, which we denote by decision time step. This reflects the fact that the decision making process can last more than $\Delta t$ and decisions in a real-time framework would not be updated every $\Delta t$ \cite{Dandl2019RealTimeGaming}.

\subsection{Customer Model}
A customer’s request (e.g. made by smartphone app) is initially defined by three values: an origin $o_r \in N_a$, where a trip is supposed to start, a destination $d_r \in N_a$, where a trip is supposed to end, and implicitly the time $t_r$ when the request is sent. Directly inferred by these parameters is the customer’s direct route, which we define as the shortest path connecting $o_r$ and $d_r$ on $G$ with respect to travel time. Moreover, $t_r^{direct}$ and $d_r^{direct}$ are the travel time and the distance for this route, respectively.
\par
In this study, customers allow a certain maximum detour time. We assume that the maximum detour time $\delta$ is related to the travel time of the direct route $t_r^{direct}$ from $o_r$ to $d_r$: $\delta = \Delta detour \cdot t_r^{direct}$. Additionally, we assume that a customer is only willing to wait for a pick-up a certain amount of time $t_{wait}^{max}$. Therefore, a vehicle has to reach this customer before $t_{lp} = t_r + t_{wait}^{max}$. Similar to Boesch et al.~\cite{Boesch2016FleetSize}, we assume that a customer will already make the request before being available for pick-up at the origin $o_r$, hence she will need a minimum waiting time of $t_{wait}^{min}$ to be ready. Therefore, a vehicle has to wait until the earliest possible pick-up time $t_{ep} = t_r+t_{wait}^{min}$ if it arrives at $o_r$ before that.
\par
\begin{figure}[tb]
\centerline{\includegraphics[width=0.5\textwidth]{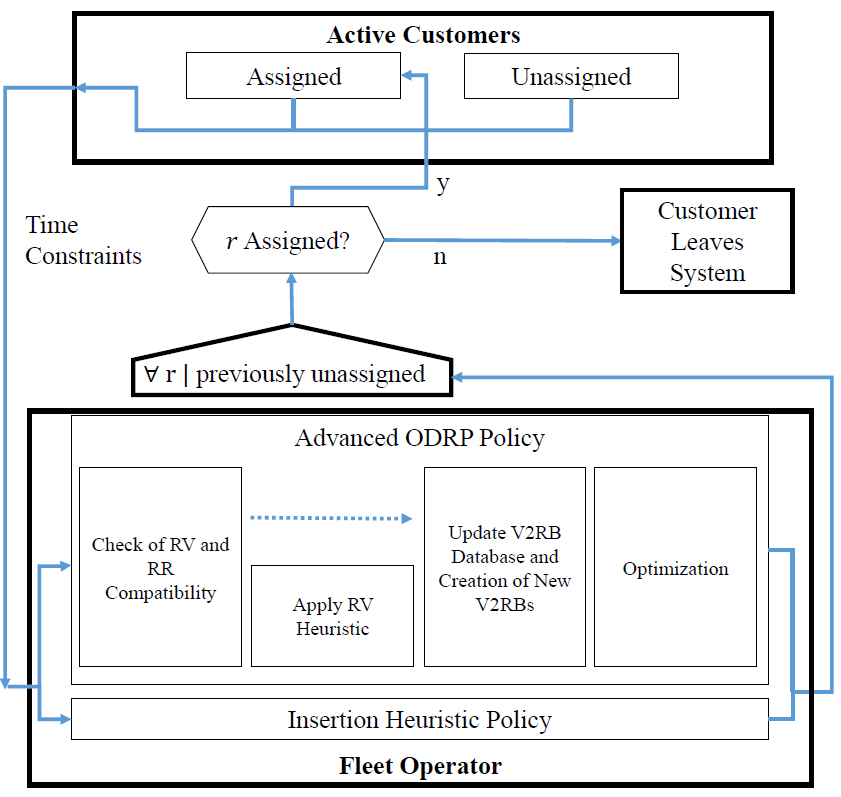}}
\caption{Flowchart describing the agent-based model with the different fleet operator policies.}
\label{overview_flowchart}
\end{figure}

The information process is also an important service quality feature, which impacts the possibility of assignment re-optimization. We assume that a customer will accept a service offer in short notice, which we define as one decision time step. If no vehicle is assigned to serve this request after this decision time step, the customer will leave the system. This behavior is shown in the Flowchart of Fig. \ref{overview_flowchart}.

\subsection{Fleet Operator Policy}
The fleet operator's task is to use its vehicles with capacity $\mu$ to maximize an objective or a set of objectives. A policy is a set of rules that the fleet operator uses to achieve these objectives. These rules dictate how customers are added to tours and which tours are assigned to vehicles. The objective in this study is to serve as many customers as possible while finding assignments that maximize ``saved VMT", which for a single tour $\xi$ we define as the difference between the sum of the direct route distances $d_r^{direct}$ of all requests $r$ involved in a tour, and the distance of the tour $d_\xi$:
\begin{equation}
U[\xi] = \left(\sum_{r \in \xi} d_r^{direct} \right) - d_\xi
\label{obj_funct}
\end{equation}
In this work, the policy dictates that the operator just allows \textit{feasible} tours, which on the on hand satisfy the time constraints of all requests, which are part of that tour, and on the other hand the number of customers on board never exceed the vehicle capacity $\mu$. This objective function is different from~\cite{alonsomora}, where the sum of detour and waiting time of users was minimized, which can prioritize driving users exclusively in case of vehicle surplus.
\par
We implement two policies in this study: A basic, but very fast insertion heuristic and an advanced ODRP algorithm based on~\cite{alonsomora}. We will explain the simpler insertion heuristic before outlining the more complex algorithm and its benefits. Lastly, we introduce a heuristic for the advanced ODRP algorithm to decrease computational effort while keeping the performance as high as possible. The flowchart for the simulation framework is shown in Fig. \ref{overview_flowchart}

\subsubsection{Insertion Heuristic Policy}
In this policy, the fleet operator processes all unassigned requests sequentially. For every unassigned request $r$ and every vehicle $v_x$, new possible tours are created by inserting stops for boarding at $o_r$ and leaving at $d_r$ into the tour currently assigned to the vehicle. The fleet operator checks all possible permutations for feasibility and the utility function $U$ defined in Eqn.~\ref{obj_funct} is computed for all feasible tours. For each vehicle $v_x$, the utility function of all feasible tours including request $r$, denoted by $\{\xi_r^x\}$, are compared with the utility of the original tour $\xi_0^x$ (before including request $r$):
\begin{equation}
\Delta U^x = \max_{\xi \in \{\xi_r^x\}} (U[\xi] - U[\xi_0^r])
\label{eq_ih}
\end{equation}
The operator designates the vehicle $v_x$ with the highest increase in utility $\Delta U^x$ to serve request $r$ by assigning the tour corresponding to Eqn.~\ref{eq_ih}.
\par
The weakness of this insertion heuristic is that it neglects benefits from global optimization and re-assignments. Instead of trying to optimize the system globally, the insertion heuristic assigns requests sequentially. However, dynamic systems, like the agent-based model used in this study, benefit from dynamic solutions with re-assignments. Fixing the assignments, as the heuristic does, restrains the solution space and prevents this algorithm from finding better solutions. 

\subsubsection{Advanced ODRP Re-Optimization Policy}
A possible option to overcome the limitations of the insertion heuristic is to generate all feasible tours for each vehicle and assign the best ones by global optimization. Unlike the insertion heuristic, this approach does not induce an order and re-optimization may change previous assignments in case a better solution is found. On the flipside, even with pre-processed routes between all access points, the task of creating and updating all feasible routes in real-time is troublesome because of the curse of dimensionality. In an exhaustive search algorithm, all possible tours, i.e. every possible permutation of stops for every possible combination of requests for every single vehicle, are created and checked for feasibility. For this reason, this approach cannot scale and the limit for real-time operability is reached for very small problem instances.
\par
Our policy is based on the work of Alonso-Mora et al.~\cite{alonsomora}, who developed a multi-stage graph-based approach in order to reduce the amount of explicit feasibility checks. To elaborate the key points that allow an incredible speed up compared to an exhaustive search procedure, we define an object called Vehicle-To-Request-Bundle (V2RB). A V2RB $\Psi(v_x |r_{i1},r_{i2},r_{i3},…,r_{in})$ contains all feasible tours, i.e. feasible permutations of pick-up and drop-off locations (and routes connecting these locations) of vehicle $v_x$, that serve exactly the set of requests $\{r_{i1},r_{i2},r_{i3},…,r_{in}\}$ named a request-bundle. We call a V2RB of grade $n$ if this bundle contains exactly $n$ requests.
\par
Since the number of possible stop combinations grows exponentially with $n$, reducing the amount of V2RBs and stop combinations for a V2RB that need to be checked, will lead to great savings in computational effort compared to an exhaustive search. Therefore, there are mainly four key points to check before building a new V2RB:
\begin{enumerate}
\item Request-to-Vehicle (RV) compatibility: If the vehicle $v_x$ cannot reach the requests origin before the latest pick-up time associated with this request, a V2RB $\Psi(v_x |r_{i1},r_{i2},r_{i3},…,r_{in})$ containing vehicle $v_x$ and request $r_{i1}$ is not built.
\item Request-to-Request (RR) compatibility: If there is no feasible tour connecting the origins and destinations of request $r_{i1}$ and request $r_{i2}$, a V2RB $\Psi(v_x |r_{i1},r_{i2},r_{i3},…,r_{in})$ containing these requests is not built for any vehicle $v_x$. 
\item A V2RB $\Psi(v_x |r_{i1},r_{i2},r_{i3},…,r_{in})$ of grade $n$ cannot be built before building VBRBs of lower grades containing a subset of the request-bundle. E.g. if the V2RB $\Psi(v_x |r_1,r_2)$, i.e. feasible tours serving request $r_1$ and request $r_2$ within their time windows does not exist, there is no possibility to find feasible tours for $r_1$,$r_2$ and $r_3$ (The V2RB $\Psi(v_x |r_1,r_2,r_3)$ cannot be feasible, too). Additionally, building the higher grade V2RB for a new request $r_x$ upon the corresponding lower one (without that request) saves large amount of stop combinations need to be checked: Instead of checking all possible permutations of stops, only insertions of $o_{r_x}$ and $d_{r_x}$ into the list of feasible tours of the lower V2RB have to be considered.
\item All on-board requests have to be included in V2RBs $\Psi(v_x |r_{i1},r_{i2},r_{i3},…,r_{in})$ of a vehicle.
\end{enumerate}
The key difference of our approach with that of Alonso-Mora et al.~\cite{alonsomora} is that they rebuild all these objects from scratch in every single decision time step, while we keep the computed V2RBs stored in a database in order to reduce the number of needed checks in the next decision time step. Our idea is that it is not necessary to check all possible permutations of stops again; instead, we only check if the previously feasible tours in the V2RB $\Psi(v_x |r_{i1},r_{i2},r_{i3},…,r_{in})$ are still feasible. Between decision time steps, the status of a previously feasible tour can change as vehicles could move or customers could board or leave a vehicle. For example, as a vehicle $v_x$ moves along its assigned tour in direction of stop $A$, there can be tours in the V2RBs $\Psi(v_x |r_{i1},r_{i2},r_{i3},…,r_{in})$ of this vehicle, where another first stop $B$ cannot be reached in time anymore. Likewise, there is no point of keeping possible tours that do not contain the customer, who has just boarded the vehicle.
\par
While \cite{samitha2019} also keep already computed RV and RR compatibilities in memory, we also store all computed V2RBs containing all feasible stop combinations. Thereby, in most cases, the order of the stops in feasible routes don't have to be recomputed again in the next optimization step.
\par
Summarizing, the fleet operator conducts following steps one-after-another in each decision time step in order to find all feasible V2RBs: (i) compute RV and RR compatibility for new requests (key point 1 and 2); (ii) update existing V2RBs; for unassigned requests (iii) compute new V2RBs from low to high grades based on existing ones following key point 3. Even by using this approach the computation time is exceeding the real time, also for low levels of demand. Therefore, computing V2RBs for different vehicles on different processor cores in parallel, can decrease the computational time by magnitudes.
\par
After that, we perform a global optimization process in order to find the system’s optimal assignments in each decision time step. At this stage, each of the possible tours of a V2RB is rated by the utility function defined in equation (\ref{obj_funct}) and each V2RB itself is rated by the best utility value of its tours and represented by this tour.
\par
In this study, the objective of the fleet operator is to maximize the sum of these utility values representing ``saved VMT" of all assigned tours, while serving as many requests as possible. This optimization problem can be formulated as an Integer Linear Problem in the following way:
\begin{align}
\max_{z_{jk}} ~ ~ & \sum_{j,k} u_{jk} z_{jk} - \sum_j P x_i \label{opt_obj}\\
\text{s.t.} ~ ~ & \sum_{k \in K(i)} \sum_j z_{jk} + x_i = 1 ~ ~ \forall i | r_i \in R_{u} \label{opt_c1}\\
 & \sum_{k \in K(i)} \sum_j z_{jk} = 1 ~ ~ \forall i | r_i \in R_a \label{opt_c2}\\
 & \sum_k z_{jk} \leq 1 ~ ~ \forall j \label{opt_c3}\\
 & z_{jk}, x_i \in \{0,1\} ~ ~ \forall k,j,i \label{opt_c4}
\end{align}
Eqn.~\ref{opt_obj} represents the global objective function. $u_{jk}$ is the utility function value of the V2RB $\Psi(j|k)$, i.e. the best tour of vehicle $j$ to serve the request bundle $k$. $z_{jk} \in \{0,1\}$ is a decision variable taking the value $1$ if this V2RB is assigned and $0$ otherwise. $P$ is a large penalty term and the decision variable $x_i \in \{0,1\}$ refers to a previously not assigned request $i$. According to Eqn.~\ref{opt_c1},  $x_i$ takes the value of $1$ if $i$, contained in the set of unassigned requests $R_u$, is not assigned in this optimization. Furthermore, Eqn.~\ref{opt_c1} assures that only one V2RB $k$ in $K(i)$ (the set of all V2RBs containing request $i$) can be assigned to a vehicle $j$. A very large value of $P$ prioritizes serving requests over saving VMT. Hence, the fleet operator tries to maximize the number of assigned customers.  Eqn.~\ref{opt_c2} constrains the assignment of already assigned requests, contained in the set $R_a$, and the equality assures that previously assigned requests must be assigned again. Nevertheless, the corresponding V2RB of a request may change over time. Finally, the constraint of Eqn.~\ref{opt_c3} assures that for each vehicle $j$ not more than one V2RB $k$ can be assigned.\\
Because of the constraints in Eqn. \ref{opt_c1} $x_i$ is no independent variable. As also shown in \cite{samitha2019} it can be absorbed by the cost function values $u_{jk}$ for better implementation.

\subsubsection{Vehicle Selection Heuristic for Advanced ODRP Re-Optimization Policy}
Although the proposed multi-stage algorithm reduces the number of routes to be checked compared to an exhaustive search by magnitudes, the exponential scaling of the problem cannot be overcome. As a result, computation time of new assignments exceeds real time for big problem sizes (around 3000 vehicles with 5000 active requests to match in our case). Especially for real-world applications this is a problem that needs to be solved. In this study we introduce a heuristic by prefiltering RV-matches to decrease computational time needed for a new assignment while keeping the solution as close to optimality as possible.  We tackle this problem by using only a subset of vehicles theoretically able to serve a request defined by RV-compatibility based on three heuristic rules. These rules aim to select only the most probable vehicles to be assigned to serve the request and therefore may reduce the search space for building new V2RBs drastically and additionally reduce the solution space for the optimization step.
\par
All rules are computed for each new request sequentially while the order of these requests is random. For each request $r_i$ the set of possible vehicles defined by RV-compatibility $V_i$ is divided in the subsets $V_{i,a}$ and $V_{i,u}$, the vehicles with a current assignment and idle vehicles, respectively. With the following three rules only a subset $V_{i,build}$ of these vehicles is chosen for creating the V2RBs.
\par
Rule I: The first rule constrains the number of possible vehicles from $V_{i,a}$ for request $r_i$ to $\chi_a \in \mathbb{N}$. The decision to add a vehicle $v_j$ to the list of possible vehicles $V_{i,build}$ for request $r_i$ is based on the function
\begin{align}
\nonumber
f(v_j|r_i) = \frac{m_{r}}{m_{rr}} \min_{k,l} (tt(x_{j,l},o_{r_i}) + tt(d_{r_i}, x_{j,k}) ) \\
\label{eq:rule1}
 ~|~  m_{rr} \neq 0 ~,~ \forall v_i \in V_{i,a}
\end{align}
to measure the compatibility of the request $r_i$ to the vehicle's $v_j$ currently assigned tour. $x_{j,m}$ are the ordered stops of the currently assigned tour including the current vehicle position. The function $tt(a,b)$ measures the travel time between location $a$ and $b$. While the second factor of equation (\ref{eq:rule1}) measures the distance of the new request's origin and destination to the assigned route, the first factor measures the compatibility of the new request with the requests corresponding to the currently assigned route. $m_r$ is the number of the currently assigned requests to the vehicle and $m_{rr}$ is the count of these requests compatible with the new request based on RR. In case the new request ist incompatible with a request currently on board of the vehicle or $m_{rr} = 0$ and $m_r > 1$, this vehicle will not be considered any further. If $m_{rr} = 0$ and $m_r = 1$ the vehicle will be considered as idle and therefore added to $V_{i,u}$ for the next heuristic rules (a reassignment of vehicles with currently only one assigned request might nevertheless be favorable). The first $\chi_a$ vehicles sorted in ascending order of $f(v)$ are chosen to be possible vehicles for this request and added to $V_{i,build}$.
\par
For Rule II and Rule III each vehicle is initialized with the attributes $\vec{v_{j}}$ with $|\vec{v_{j}}| = 1$, a random 2d-vector with length one, $N_{d,j} = 0$, the number of requests added to the vehicle $v_j$ by Rule II, and $N_{n,j} = 0$, the number of requests added to the vehicle $v_j$ by Rule III.
\par
Rule II: The second rule is used to match requests with similar travel directions into the same idle vehicles from $V_{i,u}$. The normalized vector of travel $\vec{od_i}$ with $|\vec{od_{i}}| = 1$ of the new request $r_i$ is pointing into the direction of request origin $o_i$ to the request destination $d_i$. Rule II comprises the following steps:
\begin{enumerate}
\item Calculate the scalar product:
	\begin{align}
	s_{j,i} = \vec{v_{j}} \cdot \vec{od_i} ~,~ \forall v_i \in V_{i,u}
	\end{align}
\item Pick $\chi_{u,d} \in \mathbb{N}$ vehicles with highest $s_{i,j}$, add them to $V_{i,build}$ and remove them from $V_{i,u}$.
\item For each of these vehicles, update the vehicle attributes
	\begin{align}
	N_{d,j} &\leftarrow N_{d,j} +1 \\
	\label{eq:veh_vec}
	\vec{v_{j}} &\leftarrow \frac{N_{d,j}}{N_{d,j} + 1} \vec{v_{j}} + \frac{1}{N_{d,j} + 1} \vec{od_i}
	\end{align}
\end{enumerate}
By this rule the vehicle vector tends to point in the average direction of the added requests and therefore enhances clustering of new requests with similar travel directions. The idea behind this clustering is, that requests that can possibly be pooled will be added to the same vehicles and thereby enable higher V2RBs. A random initialized vehicle vector is used to not induce any specific directions for the idle vehicles and for the resulting clusters. 
\par 
Rule III: The third rule is used to match requests to the $\chi_{u,n} \in \mathbb{N}$ nearest idle vehicles with an additional factor taking into account the number of already added requests to a specific vehicle to achieve an equal distribution of requests among idle vehicles. For request $r_i$ with origin $o_i$ following steps are computed:
\begin{enumerate}
\item Calculate the distance to vehicle $v_j$ with current position $loc(v_j)$ weighted by the number of already added requests to this vehicle $N_{n,j}$:
	\begin{align}
	t_{i,j} = (N_{n,j} + 1) \cdot tt(loc(v_j), o_i) ~,~ \forall v_i \in V_{i,u}
	\end{align}
\item Pick $\chi_{u,n} \in \mathbb{N}$ vehicles with smallest $t_{i,j}$, add them to $V_{i,build}$ and remove the from $V_{i,u}$.
\item For these vehicles increase $N_{n,j}$ by 1.
\end{enumerate}
The factor $N_{n,j}$ thereby is used to achieve a homogenous distribution of request among vehicles to achieve a high service rate, the main objective of the operator in this study.
\par
The used heuristic will be characterized by the tuple $(\chi_a, \chi_{u,d}, \chi_{u,n})$ in the following, defining the number of vehicles per request picked by the introduced heuristic rules. With this heuristic, possible V2RBs for a maximum of $\chi_a + \chi_{u,d} + \chi_{u,n}$ vehicles are computed per request.

\section{Case Study in Munich, Germany}
\label{sec:caseStudy}
In the following, the described methodology is applied to the network of Munich. The network graph and the customer requests for the agent-based simulations are firstly introduced, then the parameters of the scenarios are summarized. Subsequently, results of the simulations are illustrated and discussed. 

\subsection{Inputs from Traffic Model of Munich}

\begin{figure}[tb]
\centerline{\includegraphics[width=0.4\textwidth]{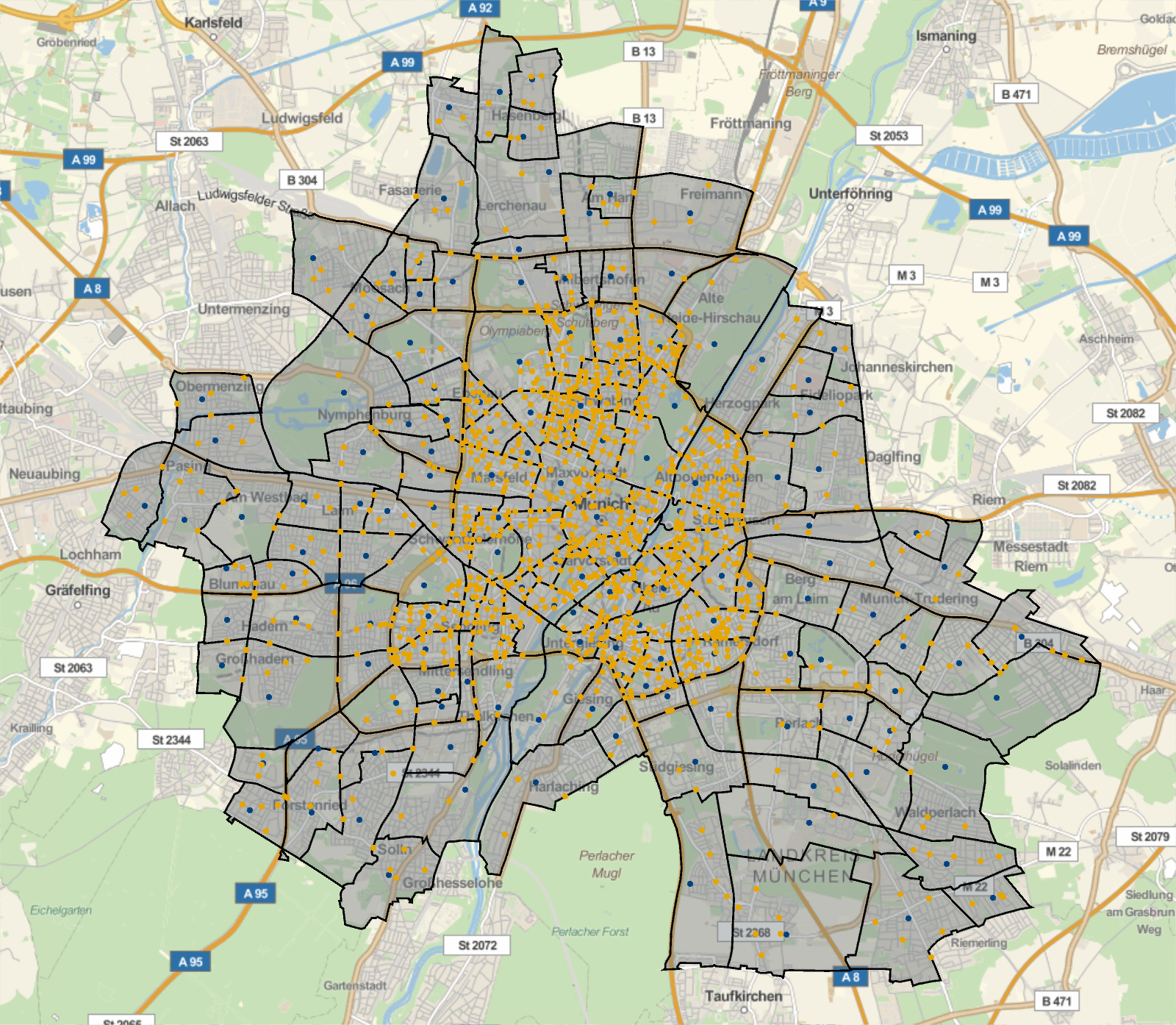}}
\caption{Map of Munich with operating area of the ride-pooling service highlighted in blue and access points in yellow. The operating area covers a surface of approximately 220 km$^2$.}
\label{operating_area}
\end{figure}

\begin{figure*}[tb]
\centerline{
a)~\vtop{\vskip-1ex\hbox{\includegraphics[width=0.35\textwidth]{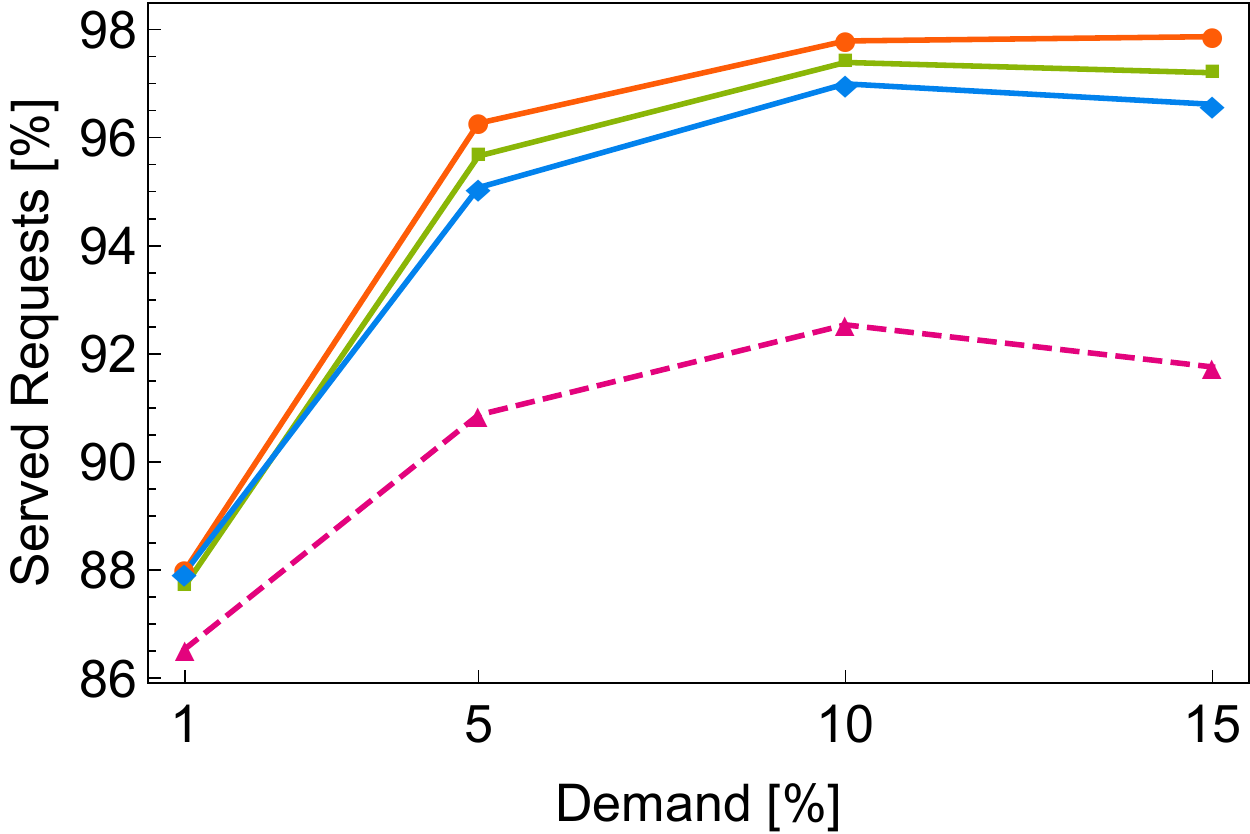}}}
~ ~
~\vtop{\vskip-1ex\hbox{\includegraphics[width=0.18\textwidth]{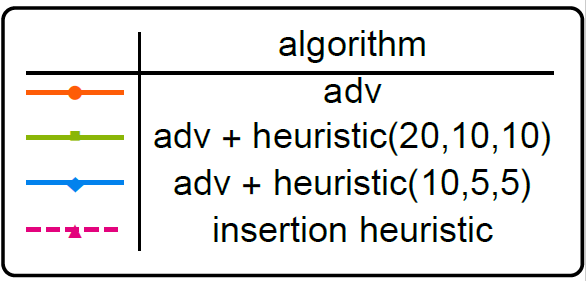}}}
~ ~
b)~\vtop{\vskip-1ex\hbox{\includegraphics[width=0.36\textwidth]{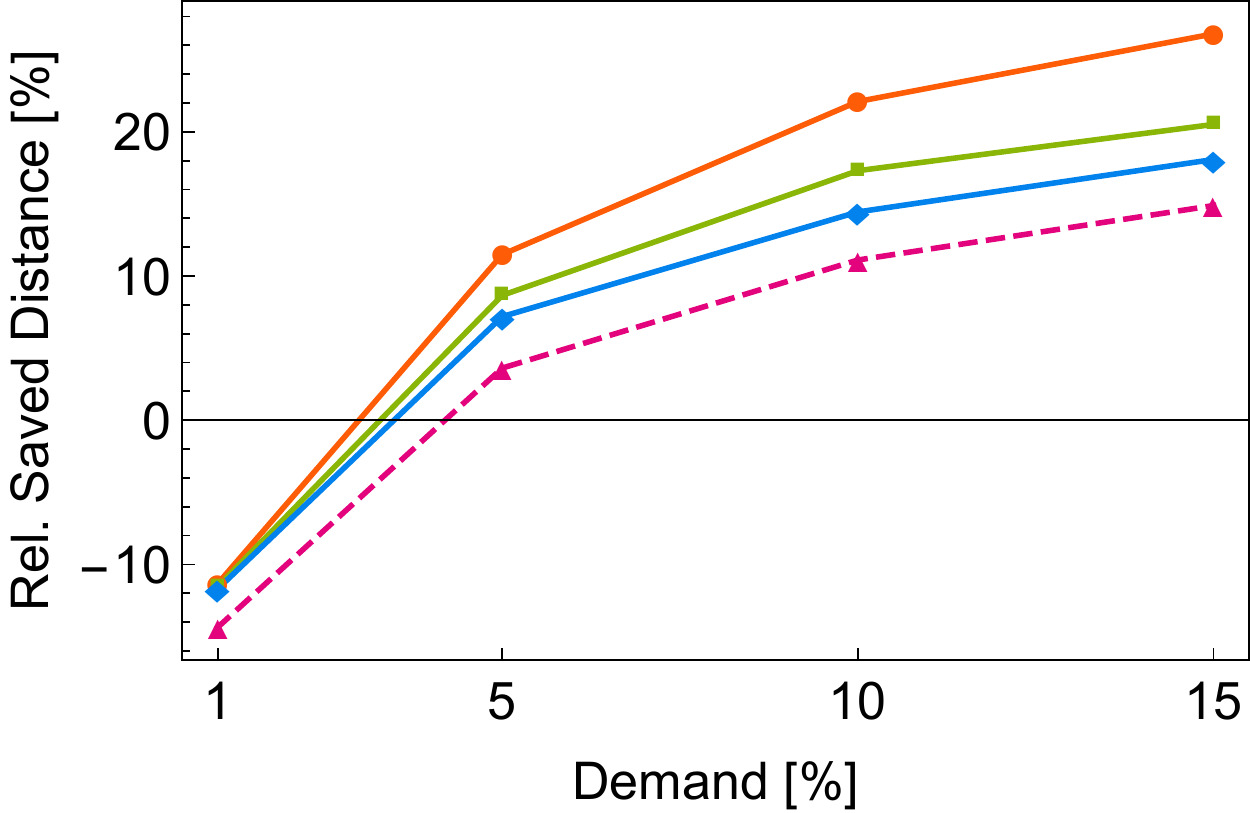}}}
}
\caption{Main evaluation criteria of scenarios for different matching algorithms and demand densities. ``adv'' corresponds to the advanced ODRP algorithm. If a heuristic is used, it is characterized by the tuple  $(\chi_a, \chi_{u,d}, \chi_{u,n})$. A demand density of 10\% corresponds to a total daily trip number of approximately 120k trips, which are served by 2000 vehicles. a) Share of served user requests. b) Relative saved distance compared to every customer's direct distance.}
\label{fig:ihcompare}
\end{figure*}

The street network $G$, as well as hourly changing travel times are extracted from a traffic micro-simulation model described in~\cite{FloBene17}. The network consists of 19522 nodes and 40952 edges. The travel times for each edge $e \in E$ are used for both forecasting route travel times, as well as vehicle movements. Shortest paths between customer access points are preprocessed. All other shortest paths are calculated during the simulation using Dijkstra's algorithm with preprocessing techniques from \cite{Delling2009EngineeringRP}.
\par 
A share $D$ of private vehicle trips within an operating area are the basis for the demand of the ODM system. The determination of the operating area is demand driven in this study. An initial analysis of private vehicle origin-destination (OD) relations showed an increasing trip density the closer a district is located to Munich's city center and a compact operating area which is shown in Fig.~\ref{operating_area} is defined.
\par
Altogether, the private vehicle OD matrices count 1.2 million trips per day between 1423 access points within the resulting operating area partitioned into 15-minute slices. We vary demand by globally modifying the adoption rate $D$, the share of private vehicle trips replaced by the ride-pooling system. The requests for the ODRP system are created by Poisson processes, whereas the mean of the Poisson distribution corresponds to $D \cdot \lambda^{OD}$ for each origin-destination pair (with original matrix entry $\lambda^{OD}$) within the operating area, for each of the 15-minute time slices.

\subsection{Scenario Setup}
The constant as well as the variable parameters of the agent-based simulations are summarized in TABLE~\ref{scenario_setup_table}. To explore the impact of demand density, the shares of 1\%, 5\%, 10\% and 15\% from the total private vehicle OD trips within the operating area are converted to requests for the ODRP system. To extract the scaling property, we coupled fleet size linearly to demand in this study's scenarios. We started with a fleet size of 200 vehicles for 1\% demand and added 200 vehicles to the fleet for every 1\% increase in demand. This value has shown to be a good tradeoff between fleet utilization and number of served requests in test simulations. Additionally, we also considered a boarding time of 30 seconds for every stop at access points. We compare the performance of the advanced ODRP algorithm without heuristic, the ODRP algorithm with various proposed heuristic values and the insertion heuristic policy.
\par
Simulations are performed on a 3400 MHz Computer with 16 cores. Simulations with demand density 1\%, 5\%, 10\% and 15\% are computed on 2, 4, 8 and 16 cores in parallel, respectively. Gurobi Solver is used for solving the ILP optimization problem with a timeout set to 20s.

\begin{table}[tb]
\caption{Parameters of ODRP Simulations}
\label{scenario_setup_table}
\centering
\begin{tabular}{|p{3.1cm}|c|r|l|}
\hline
\textbf{Constant Parameter}                      & \textbf{Math Notation} & \textbf{Value}  & \textbf{Unit} \\ \hline
simulation time step                             & $\Delta t$             & $1$             & second        \\ \hline
time between decision time steps                 & $\Delta t^{dec}$       & $30$            & second        \\ \hline
boarding time                                    & $T^B$                  & $30$            & second        \\ \hline
minimal waiting time                             & $t_{wait}^{min}$       & $2$             & minute        \\ \hline
maximal waiting time                             & $t_{wait}^{max}$       & $8$            & minute        \\ \hline
acceptable detour ratio                          & $\Delta detour$        & $40$      & \%            \\ \hline
vehicle capacity                                 & $\mu$                  & $4$             & pax           \\ \hline
penalty for not assigned request                 & $P$                    & $40000$         & km            \\ \hline
                                                 &                        &                 &               \\ \hline
\textbf{Variable Parameter}                      & \textbf{Math Notation} & \textbf{Values} & \textbf{Unit} \\ \hline
penetration/adoption rate / simulated demand     & $D$                    & $1,5,10,15$     & \%            \\ \hline
fleet size                                       & $m$                    & $200 \cdot D$   & vehicle       \\ \hline
\end{tabular}
\end{table}

\subsection{Results}

\begin{figure*}[htb]
\centerline{
a)~\vtop{\vskip-1ex\hbox{\includegraphics[width=0.37\textwidth]{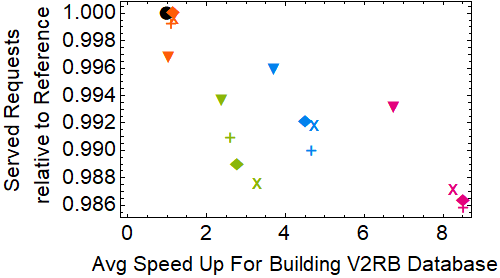}}}
~ 
~\vtop{\vskip-1ex\hbox{\includegraphics[width=0.2\textwidth]{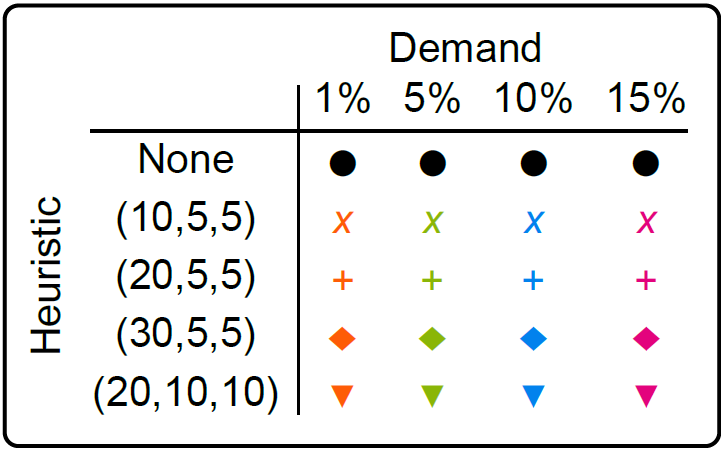}}}
~ 
b)~\vtop{\vskip-1ex\hbox{\includegraphics[width=0.37\textwidth]{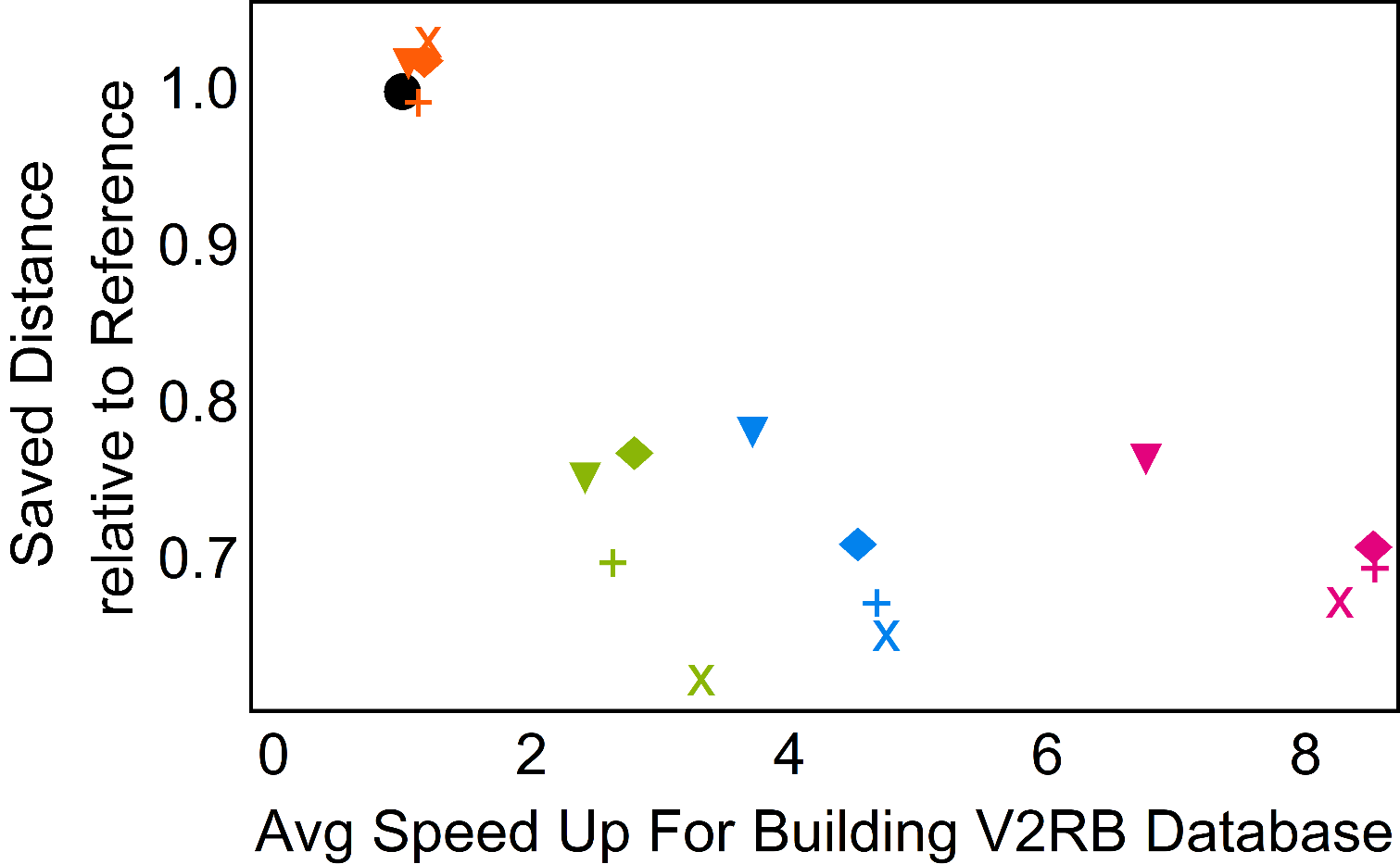}}}
}
\caption{Speed Up for building the V2RB Database vs Performance of various heuristics. The average computational time throughout the whole simulation is measured and compared. Reference scenario corresponds to the scenario without heuristic and same demand level. }
\label{fig:perf_heu}
\end{figure*}

To examine the performance of the proposed algorithms, all scenarios are evaluated globally based on the optimization objectives, number of served requests on the one hand and ``saved distance'' on the other hand. For better comparison between the different demand levels, ``saved distance is normalized to ``relative saved distance" $rsd$, defined as
\begin{equation}
rsd = \frac{\left(\sum_r d_r^{direct}\right) - \sum_v d_v}{\sum_r d_r^{direct}}~,
\end{equation}
where $d_r^{direct}$ is the direct route distance of the served request $r$ and $d_v$ is the total driven distances of vehicle $v$ in the fleet. If this value is negative, a larger distance is driven by the system's vehicles compared to the case where all the served requests would drive on their own the fastest route due to empty pick-up trips. For positive values, pooling of multiple users overcomes this extra mileage. 

\subsubsection{Comparison of advanced ODRP algorithm with insertion heuristic policy}

Fig. \ref{fig:ihcompare} shows simulation results when using advanced ODRP algorithm, the advanced ODRP algorithm with two different heuristics and the insertion heuristic. The optimization objectives ``Served Requests'' and ``Relative Saved Distance'' are plotted for different demand levels.
\par
Comparing the fraction of requests that could be served, it is evident, that the advanced ODRP algorithm performs way better than the simple insertion heuristic. The main reason is that the insertion heuristic doesn't benefit from re-optimizations due to fixed assignments. The advanced algorithm however can reassign customers that haven't boarded the vehicle and can therefore react much more flexible to dynamically incoming demand. Using the advanced algorithm, an operator can therefore raise the number of served requests by around 7$\%$ at 15$\%$ demand in our case.
\par
Comparing the performance on the objective ``relative saved distance'', the advanced algorithm still performs better than the insertion heuristic, but the difference is not that distinct compared to the objective ``served requests''. Due to the large penalty value for unassigned requests in the optimization problem (Eqn. \ref{opt_obj}) serving as much customers as possible is the prioritized objective on which the advanced algorithm clearly performs better. Of course, these additional served requests have influence on the objective ``relative saved distance''.
\par
This evaluation gives a strong argument for fleet operators to use a good matching algorithm. The differences in performances are directly correlated to the revenues an operator receives and the influence the mobility service has on the street network.

\begin{figure*}[tb]
\centerline{
a)~\vtop{\vskip-1ex\hbox{\includegraphics[width=0.39\textwidth]{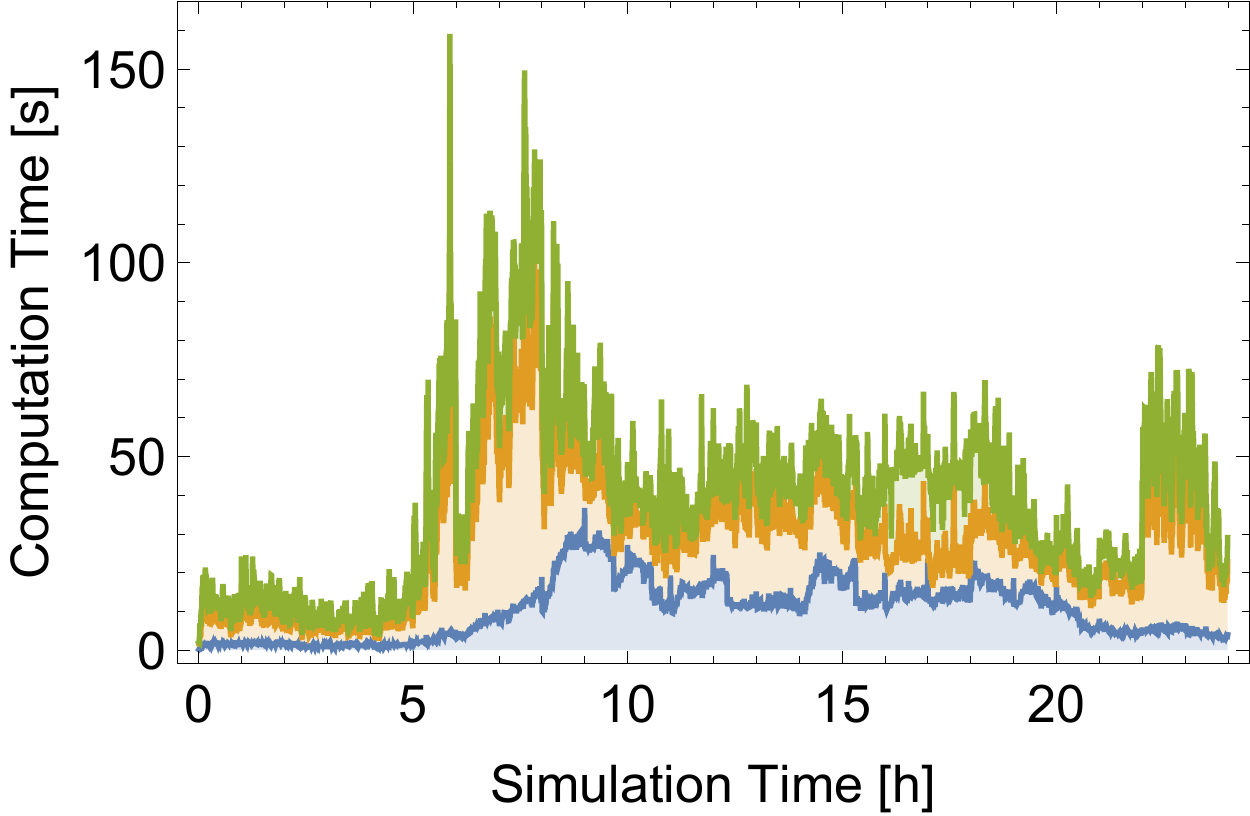}}}
~ 
~\vtop{\vskip-1ex\hbox{\includegraphics[width=0.16\textwidth]{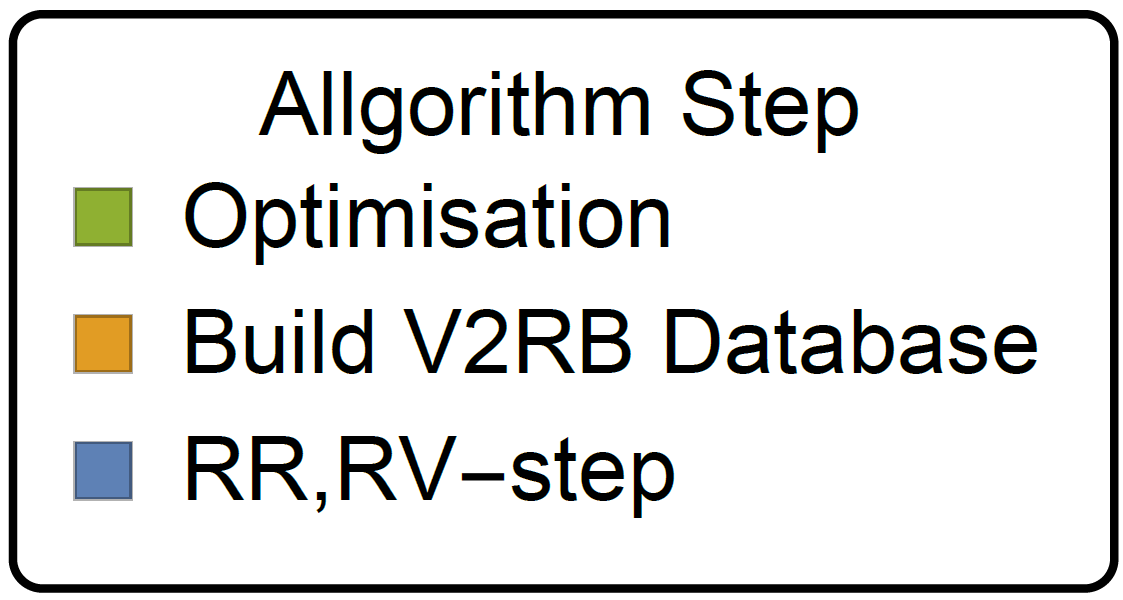}}}
~ 
b)~\vtop{\vskip-1ex\hbox{\includegraphics[width=0.39\textwidth]{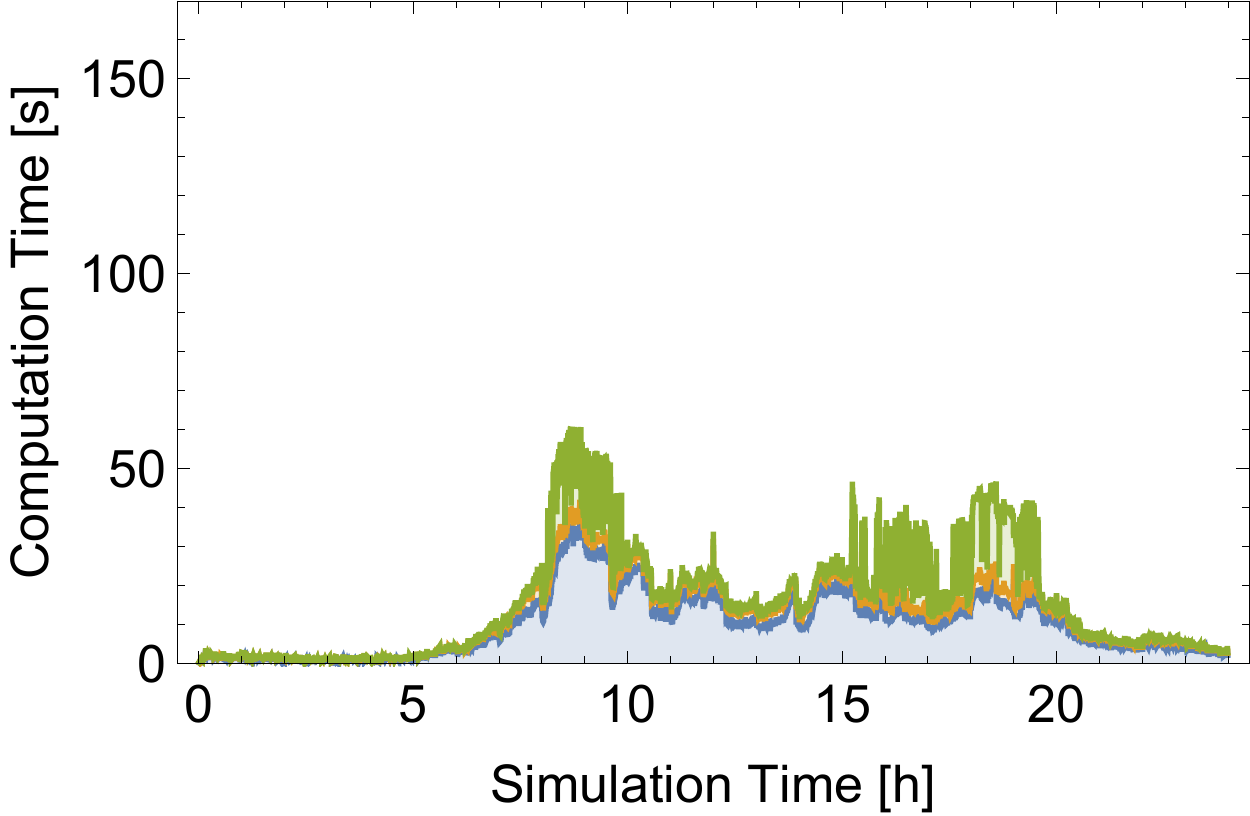}}}
}
\caption{Stacked plot of the computational time of the most costly algorithm steps at each decision time step. a) 15$\%$ scenario without heuristic. b) 15$\%$ scenario with heuristic $(20,5,5)$. }
\label{fig:comp_time}
\end{figure*}

\subsubsection{Performance of heuristic for ODRP algorithm}
Fig. \ref{fig:ihcompare} also shows the performance of two different heuristics, namely $(\chi_a, \chi_{u,d}, \chi_{u,n}) = (10,5,5)$ and $(\chi_a, \chi_{u,d}, \chi_{u,n}) = (20,10,10)$ are shown, with parameters described in the previous section. The sum of these values corresponds to the maximal number of vehicles that are considered in each optimization time step for each incoming request. Therefore, roughly speaking, the higher this number the closer the heuristic is to the exact algorithm because more vehicles and their designated requests are considered in the optimization time step until all possibilities are fully exploited. This results in the convergence of the simulations with heuristic to those simulations without for low levels of demand (1$\%$). Due to small vehicle and request density during the simulation, the heuristic hardly constrains the vehicle selection. However, even in simulations with high demand levels, nearly as much requests could be served compared to scenarios without heuristic. Serving requests is the primary objective in Eqn. \ref{opt_obj} due to the large value of $P$. The very good service rates of simulations with heuristic show that requests are being distributed among the vehicles resulting in their assignment. On the other hand, the reduction of ``saved distance'' is more prominent  due to the diminished solution space available for the optimization step.
\par

Fig. \ref{fig:perf_heu} compares the objectives and the speed-up achieved for various heuristics. Values are depicted relative to the simulation with no heuristic used at the same demand level. The speed-up only measures the computational time of building the V2RB database in the described multistage advanced ODRP algorithm. This step is most time consuming for large problem sizes and it is the step the heuristic directly influences. Again, the heuristic has nearly no influence for low demand scenarios, because the problem size is too small. For all parameters $(\chi_a, \chi_{u,d}, \chi_{u,n})$ tested, the number of served requests is influenced only minimally. Only around $1.4\%$ less requests could be served while gaining a speed up of over $8$ in the case of the $(20,10,10)$ in the 15$\%$ scenario. In this scenario around 70$\%$ of the ``saved distance'' could be maintained with the heuristic. Interestingly, for the tested demand levels the performance of the heuristics with same parameters is not strictly decreasing with demand, although the restriction of the solution space increases. The reason on the one hand is, that the difference between optimization solutions gets statistically smaller and on the other hand the used heuristic rules, mostly rule II and rule III, are statistically more effective with bigger problem sizes. Especially the vehicle vector of rule II (Eqn. \ref{eq:veh_vec}) is converging to a specific direction, defined by the set of added requests.
\par
Measured globally, instead of just the time needed for building the V2RB database, a speed-up of around 2.5 could be achieved for the tested heuristics maximally. The heuristic is most effective during peak times with high demand density and it directly affects only the computational step, when the database for feasible V2RBs is built. In Fig. \ref{fig:comp_time} the computational times for the three most time consuming algorithm steps, testing RR-\& RV-compatibility, building the V2RB database and solving the optimization problem, are shown for a one day simulation. A 15$\%$ scenario without heuristic and with  $(20,5,5)$-heuristic is compared. It is evident that the computational step of building the V2RB database could be reduced significantly, while the RR-\& RV-step is not influenced by the heuristic at all. In these two scenarios the average computational time of RR-\& RV-step, Build V2RB Database and Optimization could be reduced from $10.3s$, $18.3$s and $10.2s$ to $10.0s$, $2.1s$ and $3.5s$ for with and without heuristic, respectively. This shows, that the heuristic also has a large impact on the optimization problem as it reduces the dimension of the solution space. Overall, a speedup of $2.5$ could be achieved. This is expected to be even higher, when computing larger problems, because the heuristic acts on the computational steps scaling exponentially with problem size, while the unaffected RR-\& RV-step just scales quadratically.

\section{Conclusion}
\label{sec:conclusion}

\subsection{Summary}
In this study we elaborated an advanced multi-step ODRP algorithm, firstly introduced by \cite{alonsomora} and compared its performance on a simulation study for Munich, Germany  with an insertion heuristic algorithm. It could be shown, that by using a more refined (but computationally more expensive) algorithm an operator could serve an additional 8$\%$ of requests, while saving an additional relative distance of 10$\%$.  This stresses the importance of using advanced pooling algorithms especially for real-world applications because of the direct impact on operators revenue, customer satisfaction and the service's impact on the traffic state.
\par
We introduced some speed-up techniques for the advanced algorithm in addition to \cite{alonsomora} and \cite{samitha2019} based on keeping already computed routing possibilities in memory in order to avoid recomputing them in each decision time step again. A heuristic based on the expectation of the most probable vehicles for serving a request was developed. Acting on the computationally most expensive steps, it leads to a speed-up of a factor of 8 in the corresponding steps and a speed-up of a factor of 2.5 overall while keeping the number of served requests almost constant and losing around 30$\%$ of the secondary objective ``saved distance''. This shows, that a refined vehicle selection heuristic can help counteracting the exponential scaling of the pooling problem to retain real time computation.

\subsection{Future Work}
Firstly, we want to address the speed-up gained by keeping computed V2RBs in memory compared to calculating them again in each time step. It is expected that there is a tradeoff between checking the feasibilities of tours feasible in the former time step, but unfeasible in the current time step, and just recomputing feasible tours which have been computed before. This will therefore depend on how long a tour stays feasible on average.
\par 
Despite showing promising results, we are convinced a more refined vehicle selection heuristic can show even better results. Finding meaningful KPI's depending on request characteristic and fleet state could help for the development. Using neural networks for a preselection of used vehicles will also be tested in future.
\par
Additionally, we want to include short-term forecasts of demand in the assignment process and a pooling specific relocation algorithm.
\par
Finally, we want to connect this framework with a traffic micro-simulator to study effects on the street network in more detail.

\section*{Acknowledgment}
Funding is provided by the German Federal Ministry of Transport and Digital Infrastructure through the project “EasyRide”.


\bibliographystyle{unsrt} 
\bibliography{bib.bib}

\end{document}